\documentclass{article}
\usepackage{amsmath}
\usepackage{amssymb}
\usepackage{amsthm}
\usepackage{graphicx} 
\usepackage{authblk}
\usepackage[letterpaper, margin=1in]{geometry}
\usepackage{natbib}
\usepackage{hyperref}
\usepackage{booktabs}
\usepackage{xcolor} 
\usepackage[normalem]{ulem} 
\DeclareRobustCommand\strike[1]{\bgroup\markoverwith{\textcolor{blue}{\rule[0.5ex]{2pt}{0.6pt}}}\ULon{#1}}
\usepackage{caption}

\title{The many faces of multivariate information: Towards a generalized theory of higher-order dependencies in complex systems}
\author[1,2]{Thomas F. Varley}
\affil[1]{Vermont Complex Systems Institute, University of Vermont, Burlington, VT, USA}
\affil[2]{Department of Computer Science, University of Vermont, Burlington, VT, USA}
\date{\today}

\newcommand{\ent}{\mathcal{H}}
\newcommand{\mi}{\mathcal{I}}
\newcommand{\tc}{\mathcal{T}}
\newcommand{\dtc}{\mathcal{D}}
\newcommand{\oinfo}{\mathcal{O}}
\newcommand{\sinfo}{\mathcal{S}}
\newcommand{\xjoint}{\mathbf{X}}
\newcommand{\yjoint}{\mathbf{Y}}
\newcommand{\kl}{\mathcal{KL}}
\newcommand{\probp}{\mathcal{P}}
\newcommand{\probq}{\mathcal{Q}}

\newtheorem{theorem}{Theorem}[section]

\begin{document}

\maketitle

\begin{abstract}
Total correlation, dual total correlation, S-information, and O-information are usually presented as a family of related but distinct measures of higher-order dependency in complex systems. 
Here we show that three of them (S-information, dual total correlation, and negative O-information) are in fact collinear samples of a single affine function of an integer parameter $k$, which we call $\Delta^{k}$. 
This form provides a well-defined notion of interaction order: fragility under single-variable deletion. 
We then show that $\Delta^{k}$ has a conjugate partner, $\Gamma^{k}$, which arranges S-information, total correlation, and O-information into a dual spectrum of redundancy rather than synergy. 
Finally, we relax the implicit maximum-entropy reference distribution built into all of these measures, defining a Kullback-Leibler generalization $\Lambda^{k}(P\|Q)$ that measures higher-order information gain when updating from an arbitrary prior to an arbitrary posterior.
\end{abstract}

\section{Introduction}

A defining feature of complex systems is the phenomenon of ``emergence": when a system displays behaviors or qualities that are not trivially reducible to the properties of its constituent parts. 
While philosophers have grappled with the problem of emergence for centuries, it is only comparatively recently that scientists and mathematicians have begun to provide rigorous, formal definitions of emergence \cite{hoel_when_2017, rosas_reconciling_2020, rosas_software_2024, varley_emergence_2022, barnett_dynamical_2021}.
Information theory has emerged as a natural framework for quantifying emergence in systems and data, as it provides a natural language with which to discuss relationships between ``wholes" and ``parts" \cite{varley_information_2025}.
Constructs like entropy and mutual information (described below) elegantly describe how one's beliefs about the behavior of a system change when considering different scales of interaction - which is one of the key features of rigorously reasoning about emergence. 

The fundamental object of study in information theory is the Shannon entropy, which quantifies the average uncertainty an observer has about the state of some system $\mathbf{X}=\{X_1,\ldots,X_N\}$ that draws states from a probability distribution $\mathcal{P}(\mathbf{x})$.
Here we will write information-theoretic terms as \textit{functionals}, which take in probability distributions and return real-valued scalars:

\begin{align}
    \ent(\probp) = -\sum_{\mathbf{x}\in\mathcal{X}}\mathcal{P}(\mathbf{x})\log\mathcal{P}(\mathbf{x})
\end{align}

Joint and marginal entropies form the building blocks of more sophisticated measures of multi-element interaction, such as the mutual information:

\begin{align}
    \mi(\probp_{1;2}) &= \ent(\probp_{1}) - \ent(\probp_{1|2})\\
    &= \ent(\probp_1)+\ent(\probp_2) - \ent(\probp_{12}) \label{eq:mi_1} \\
    &= \ent(\probp_{12}) - \big[\ent(\probp_{1|2}) + \ent(\probp_{2|1})\big] \label{eq:mi_2}
\end{align}

Here we are going to abuse the notation a bit. 
$\ent(\probp_{1|2})=\ent(X_1|X_2)$, even though technically $\probp(X_1|X_2)$ is a family of distributions, rather than a single one.
This convention keeps every information-theoretic term a functional of a single distribution object, which simplifies the notation throughout, and will be necessary when we introduce divergences that compare distributions. 

The mutual information, while very powerful, only considers pairwise dependencies between individual elements (or sets of elements). 
This is a significant limitation when studying complex systems, which can contain hundreds, or thousands of mutually-interacting parts. 
This has \strike{lead} {\color{red}led} to sustained interest in the problem of generalizing the pairwise measure to higher-order interactions \cite{rosas_quantifying_2019,varley_topology_2025}.
It turns out that, for three or more variables, there is no unique multivariate extension of bivariate mutual information - instead, several valid generalizations exist, which are generally understood as revealing different ``kinds" of dependency, and track different intuitions about what multivariate information might mean. 

\subsection{Measures of multivariate information}

\subsubsection{Total correlation}
First introduced by Watanabe \cite{watanabe_information_1960}, and later independently re-derived by Tononi, Sporns, and Edelman \cite{tononi_measure_1994}, the total correlation is the natural extension of Eq. \ref{eq:mi_1} and measures the degree to which the whole system deviates from independence:

\begin{align}
    \tc(\probp) := \bigg(\sum_{i=1}^{N}\ent(\probp_i)\bigg) - \ent(\probp) 
\end{align}

If, for all $i$, $\probp_i\bot \probp^{-i}$, then $\tc(\probp)=0$ bit, the minimum possible value. 
In contrast, if all elements are copies of each other, the total correlation achieves its maximum possible value (for given marginal entropies) of $(N-1)\ent(\probp)$.
The total correlation can also be represented as a Kullback-Leibler divergence:

\begin{equation}
    \tc(\probp)=\kl(\probp ||\prod_{i=1}^{N}\probp_i).
\end{equation}

This approach defines the total correlation as the information gained when updating one's model of a system from a prior of maximum-entropy independence to the true joint statistics.

\subsubsection{Dual total correlation}
The second-oldest measure discussed here is the dual total correlation, originally introduced by Han \cite{han_nonnegative_1978} and later reformulated by Abdallah et al., \cite{abdallah_measure_2012}. It is the natural generalization of Eq. \ref{eq:mi_2}.
Where the total correlation is a measure of global dependence, the dual total correlation measures the amount of ``shared information" in the system: how much of the joint entropy is distributed over two or more variables:

\begin{align}
    \dtc(\probp) = \ent(\probp) - \sum_{i=1}^{N}\ent(\probp_{i\mid \setminus i}) \label{eq:dtc}
\end{align}

Here, $\ent(\probp_{i\mid\setminus i})$ is notational short-hand for the residual entropy \cite{rosas_quantifying_2019}: the uncertainty in variable $i$ that persists after conditioning on all other elements $\setminus i$. The sum of residual entropies, $\sum_{i}\ent(\probp_{i\mid\setminus i})$, is also known as the ``erasure entropy" \cite{verdu2008}. 
Like the total correlation, the dual total correlation is zero in cases of global independence, but it is also low (relative to the total correlation) in cases of total synchrony as well. 
It is maximal when there is large amount of information represented by the joint states of many elements.  

\subsubsection{S-information and O-information}
James et al., explored two measures of higher-order interaction constructed from sums and differences of the total correlation and dual total correlation \cite{james_anatomy_2011}.
The first, which was later renamed the S-information by Rosas et al., \cite{rosas_quantifying_2019}, is the sum of the total correlation and dual total correlation:

\begin{align}
    \sinfo(\probp) &= \tc(\probp) + \dtc(\probp) \\
    &= \sum_{i=1}^{N}\mi(\probp_{i;\setminus i}), \label{eq:s}
\end{align}

Like the total and dual total correlation, the S-information is non-negative, and zero only when all elements are independent. 
It is maximal when each element is high entropy and strongly coupled to the rest of the system.
The S-information is often thought of as a measure of the ``total" information in the system, and as we will see, serves as a natural reference point when exploring the landscape of higher-order information measures. 

The final measure is the O-information. 
Unlike the above measures, the O-information is signed: if $\oinfo(\probp)<0$, then the structure of $\probp$ is understood to be dominated by synergistic interactions, while if $\oinfo(\probp)>0$, $\probp$ is dominated by redundant interactions. 

\begin{align}
    \oinfo(\probp) = \tc(\probp) - \dtc(\probp)
\end{align}

The O-information has become a popular measure in neuroscience \cite{varley_multivariate_2023,gatica_high-order_2021, kumar_g_changes_2024}, physiology \cite{mijatovic_assessing_2024,scagliarini_gradients_2024} and even been applied to music theory \cite{scagliarini_quantifying_2022}. 
Unlike other approaches to quantifying synergistic interactions in complex systems (e.g. the partial information decomposition \cite{williams_nonnegative_2010}), it scales gracefully with system size, does not require \textit{ad hoc} definitions of redundancy, and has efficient parametric estimators for real-valued data. 
Collectively, these measures create a foundation on which to build a rich picture of higher-order interactions \cite{rosas_characterising_2025} in terms of dependency strength and type of dependency (redundant or synergistic).

\section{The general form}

Recently, it was shown that the dual total correlation could be written in terms of sums of joint and marginal total correlations \cite{varley_multivariate_2023}:

\begin{align}
    \dtc(\probp) = (N-1)\tc(\probp) - \sum_{i=1}^{N}\tc(\probp_{\setminus i}) 
    \label{eq:dtc_from_tc}
\end{align}

For proof, see the Appendix. 

Since $\sinfo(\probp)$ $= \dtc(\probp) + \tc(\probp)$, it is possible to recover the S-information and the O-information purely in terms of the total correlation in the same fashion:

\begin{align}
    \sinfo(\probp) &= N\tc(\probp) - \sum_{i=1}^N{\tc(\probp_{\setminus i}}) \\ 
    -\oinfo(\probp) &= (N-2)\tc(\probp) - \sum_{i=1}^{N}\tc(\probp_{\setminus i})
\end{align}

Clearly, all three measures are of a common form, which we will call $\Delta^{k}(\probp)$:

\begin{align}
    \Delta^{k}(\probp) &= (N-k)\tc(\probp) - \sum_{i=1}^{N}\tc(\probp_{\setminus i}). \label{eq:delta_k} \\
    &= \sinfo(\probp)-k\tc(\probp) \label{eq:delta_k_2}
\end{align}

The three measures (S-information, dual total correlation, and (negative) O-information) are then arranged into a spectrum of increasing values of $k$:

\begin{align} 
    \sinfo(\probp) &= \Delta^{0}(\probp) \nonumber \\ 
    \dtc(\probp) &= \Delta^{1}(\probp) \nonumber \\ 
    -\oinfo(\probp) &= \Delta^{2}(\probp) \nonumber
\end{align}

A significant take-away is that all of these measures can be understood as being ``whole-minus-sum" statistics, in the vein of the original $\Phi^{WMS}$ measure \cite{balduzzi_integrated_2008}.
This helps us understand them, not as a set of distinct measures, but as different faces of the same underlying property: how much more deviation from independence is there in the ``whole" compared to the leave-one-out marginals?
The degree to which the wholes and parts contribute to the  balance is driven entirely by the parameter $k$, which scales the whole by some integer value.

\subsection{\texorpdfstring{The $\mathbf{k}$ parameter.}{The k parameter.}}
We can dissect the $\Delta^{k}$ function to understand the role of the $k$ parameter.
If a system were composed purely of interactions of order $k$, the sum of all leave-one-out marginals would be $(N-k)\tc(\probp)$; the total correlation of each dependency gets added $N-k$ times (once for each instance it survives). 
This gives us a tractable interpretation for the term $(N-k)\tc(\probp)$: if $\probp$ were composed entirely of $k^{\text{th}}$-order interactions, it is the expected sum of the total correlations of all $N$ leave-one-out marginals. 
The summation term, $\sum_{i=1}^{N}\tc(\probp_{\setminus i})$, is the actual sum of the leave-one-out marginals. 
The difference between $(N-k)\tc(\probp)$ and $\sum_{i=1}^{N}\tc(\probp_{\setminus i})$ is the deviation of the true statistics from the idealized case where all interactions are of order $k$. 
If $\Delta^{k}(\probp)>0$, then $\sum_{i=1}^{N}\tc(\probp_{\setminus i}) < (N-k)\tc(\probp)$.
This means that the actual sum of the leave-one-out marginals is decaying faster than would be expected in the case where $\probp$ was comprised entirely of interactions of order $k$. 
This occurs when dependencies whose effective orders are greater than $k$ are highly represented, since they are erased more easily by single-element failures than dependencies of order $k$ are. (We make the notion of effective order precise below, and contrast it with irreducible order in Section \ref{sec:multiple_orders}.) 
In contrast, $\Delta^{k}(\probp)<0$ when the sum of the leave-one-out marginals decays more slowly than would be expected: i.e. the dependencies are of a lower order than $k$ and survive more of the single-element removals.

This leads to a nice generalization of a known property of the O-information: that it is zero when only pairwise dependencies are present.
The $\Delta^{k}$ function generalizes this pattern: for a system $\probp$ composed of blocks of purely $k^{\text{th}}$-order dependencies, $\Delta^{k}(\probp) = 0$ (this follows from the property that $\Delta^{k}$ is additive over independent subsets of $\probp$, for proof see Proof \ref{proof:additive_delta_k}).
Note that this implication only runs one way: a pure distribution of effective order $k$ will have $\Delta^{k}=0$, however a mixed distribution may have $\Delta^{k}=0$ if lower and higher-order dependencies balance out (as is the case with the O-information). 
Section \ref{sec:lambda_camouflage} shows how an explicit reference distribution can resolve this issue.

\subsection{Crossing the zero boundary}

While $k$ is technically unbounded, it very quickly stops being meaningful in any practical sense, and since it defines a straight line, sweeping values of $k$ isn't particularly informative. 
A more interesting question is at what value of $k$ does the curve cross zero?
Intuitively, this value (which need not be an integer) gives an \textit{effective order}: the order of a pure interaction that would lose its total correlation under single-element failures at the same rate as $\probp$ does. 
If we set $\Delta^{k}(\probp) = \sinfo(\probp)-k\tc(\probp)=0$ and solve for $k$, we find that it is a straightforward ratio of dual total correlation and total correlation:

\begin{align}
    \kappa_{\Delta} &= \frac{\sinfo(\probp)}{\tc(\probp)} \\
    &= 1 + \frac{\dtc(\probp)}{\tc(\probp)}
\end{align}

Note that in the case of perfect balance, $\kappa_{\Delta}=2$; exactly the value that defines the negative O-information (which is zero in that case).
The naturally discrete analogue is given by:

\begin{align}
\label{eq:integer_kappa}
    \kappa^{*}_{\Delta} = \lfloor\kappa_{\Delta}\rfloor,
\end{align}
which is the largest integer $k$ for which $\Delta^{k}(\probp)\geq0$. 
For the pure pairwise copy, $\kappa^{*}_{\Delta}=2$, and for the three-bit parity gate, $\kappa^{*}_{\Delta}=3$.
Unlike all of the usual information-theoretic measures, $\kappa_{\Delta}$ is unitless, making it a natural approach for comparing systems with different numbers of elements.

\subsection{Multiple notions of ``order": comparison with connected information and related measures}
\label{sec:multiple_orders}

There is a long history in information theory of constructing measures of dependency that provide some sort of univariate axis of dependency specific to a given scale. 
Classic examples include the connected information \cite{schneidman_network_2003}, the cohesion \cite{reing2021}, and the Tononi-Sporns-Edelman complexity \cite{tononi_measure_1994}, while more modern ``backbone" measures include the synergistic disclosure information decomposition \cite{rosas_operational_2020} and the alpha-synergy decomposition \cite{varley_scalable_2024}. 

Being built out of the total correlation and Kullback-Leibler divergence, the connected information and the $\Delta^{k}$ have superficial similarity, and coincide in the case of the purely synergistic, N-bit parity gate. 
For an N-bit XOR distribution, $\Delta^{N}=0$ bit and $\kappa_{\Delta}=N$, just as the connected information will place all of the dependence in the $N^{\mathrm{th}}$-component. 
In general, though, the two measures define very different notions of ``order." 
$\Delta^{k}$ asks about the robustness of the global total correlation under single-element failures \textit{relative to an expected level of fragility} (fixed by $k$), while connected information asks whether lower-order marginals are sufficient to pin down the full joint. 
We can see that distinction when we consider the W-distribution (also called the one-hot distribution, see Table \ref{tab:w_dists}). 

\begin{table}[h]
\centering
\caption{The W-distribution.}
\begin{tabular}{@{}cc@{}}
\toprule
$\xjoint$   & $\probp^{W}(\mathbf{x})$ \\ \midrule
(0,0,1) & 1/3  \\
(0,1,0) & 1/3  \\
(1,0,0) & 1/3 \\ 
\bottomrule
\end{tabular}
\label{tab:w_dists}
\caption*{This distribution presents an interesting distinction between different notions of order. Using the connected information \cite{schneidman_network_2003} it is of order two, while $\kappa_{\Delta}=2.35$, highlighting different notions of higher-order information-sharing.}
\end{table}

The connected information puts all of the deviation from independence in the second order: pairwise marginals uniquely pin the joint distribution: there is no deviation from independence that requires three variables to represent. 
The $\kappa_{\Delta}\approx2.35$, however - which seems to imply a greater-than-pairwise interaction. 

The rather odd interpretation is that the global total correlation of the W-distribution decays under single-element removal as if it were a ``pure" $2.35$-element dependency (or as if it were a mixture of pure second and third-order blocks). 
Such a fractional dependency is, of course, impossible, but it follows from the fact that removing a single element $X_i$ doesn't just disrupt the high-order dependencies, but both the high \textit{and} the low-order dependencies that $X_i$ participates in. 
The relative change in magnitude (captured by the ratio $\sum_{i=1}^{N}\tc(\probp_{\setminus i})/\tc(\probp)$ is what drives $\kappa_{\Delta}$. 
This isn't telling us anything about how many elements are required to specify the joint state, \textit{per se}, but rather that under single-element failures, the W-distribution decays faster than a purely-pairwise distribution would, but is nevertheless more robust than a purely third-order distribution. 

We should, then, make a distinction between what we might call ``irreducible order", which is measured by the connected information, and ``effective" or ``marginalization" order, which is what $\Delta^{k}$/$\kappa_{\Delta}$ are sensitive to.

A related family of measures, the cohesion \cite{rosas2016,reing2021}, bears superficial similarity as it produces a one-dimensional sequence of values that describe the dependency structure of a distribution, but its $k$ indexes the size of the subsets whose entropies are compared to the whole. 
In contrast, in $\Delta^{k}$, only single-element removals are considered and $k$ weights the whole, which makes $\Delta^{k}$ signed and gives it a zero crossing ($\kappa_{\Delta}$) with no analogue in the cohesion.
Similarly, the Tononi-Sporns-Edelman complexity \cite{tononi_measure_1994} uses a $k$ paramter to sweep scales and produces a univariate ``spectrum", but like the cohesion, it is sweeping subsets of size $k$ and likewise has no sign or zero-crossing point. 

\subsection{Examples}
Here we will explore a number of examples to highlight the properties and limitations of the $\Delta^{k}$ construction. 

\paragraph{Distributions with fixed total correlation:}

\begin{table}[h]
\centering
\caption{Two distributions with the same total correlation}
\begin{tabular}{@{}ccc@{}}
\toprule
$\xjoint$   & $\probp^{\textrm{copy}}(\mathbf{x})$    & $\probp^{\textrm{parity}}(\mathbf{x})$    \\ \midrule
(0,0,0) & 1/5 & 1/5 \\
(0,0,1) & 1/5 & 1/20 \\
(0,1,0) & 1/20 & 1/20 \\
(0,1,1) & 1/20 & 1/5 \\
(1,0,0) & 1/20 & 1/20 \\
(1,0,1) & 1/20 & 1/5 \\
(1,1,0) & 1/5 & 1/5 \\
(1,1,1) & 1/5 & 1/20 \\ \bottomrule
\end{tabular}
\label{tab:delta_dists}
\end{table}

We will start with two distributions that have the same total correlation, but different order-structure (see Table \ref{tab:delta_dists}).  
\begin{enumerate}
    \item $\probp^{\mathrm{copy}}$: Purely dyadic: it places greater probability mass on states where $X_1 = X_2$ and leaves $X_3$ independent. 
    \item $\probp^{\mathrm{parity}}$: Incorporates a higher-order interaction: probability mass is placed on states where $X_1\oplus X_2\oplus X_3=0$.
\end{enumerate}
By construction, both distributions have the same total correlation: $\tc(\probp^{\textrm{copy}}) = \tc(\probp^{\textrm{parity}}) \approx 0.278072\textrm{ bit}$, but different $\Delta^{k}$ profiles. 
The behavior of $\kappa_{\Delta}$ is as expected: for the copy-weighted distribution $\kappa_{\Delta}=2$ (indicating that the highest effective order structure is pairwise), while for the XOR-weighted distribution, $\kappa_{\Delta}=3$, indicating that the highest-order structure is beyond-pairwise. 
Since both distributions have the same total correlation, the total deviation from independence is the same: what differs is how that deviation is allocated over interactions. 

\paragraph{Distributions with fixed O-information}

We can, in theory, construct pairs of distributions to fix any particular information-theoretic measure, as well as values of $\kappa_{\Delta}$.
Here, we show two more distributions, this time fixing the O-information.
Rather than having the same total deviation from independence as above, these instead have the same redundancy/synergy balance. 
The construction rules are simple:

\begin{enumerate}
    \item[] Distribution 1: A pure three-way parity (XOR) interaction,
    \begin{align}
        \probp_{1}(\mathbf{x})\propto 1 + a\big[X_1\oplus X_2 \oplus X_3 = 0\big]. \nonumber
    \end{align}
    Every pair of variables is independent in this distribution, so every leave-one-out total correlation vanishes and $\kappa_{\Delta}=3$ for every $a>0$.
    \item[] Distribution 2: A mixture of a pairwise copy interaction and a three-way parity interaction,
    \begin{align}
        \probp_{2}(\mathbf{x})\propto \big(1+\big[X_1 = X_2\big]\big)\big(1+\big[X_1\oplus X_2 \oplus X_3 = 0\big]\big). \nonumber
    \end{align}
\end{enumerate}
Solving $\oinfo(\probp_1)=\oinfo(\probp_2)$ for the free parameter gives $a\approx0.919447$.
The resulting probabilities are given in Table \ref{tab:fixed_o}.

\begin{table}[h]
\centering
\caption{Two distributions with the same O-information.}
\label{tab:fixed_o}
\begin{tabular}{@{}ccc@{}}
\toprule
$\xjoint$ & $\probp_{1}(\mathbf{x})$ & $\probp_{2}(\mathbf{x})$ \\ \midrule
(0,0,0) & 0.164367 & 4/18 \\
(0,0,1) & 0.085633 & 2/18 \\
(0,1,0) & 0.085633 & 1/18 \\
(0,1,1) & 0.164367 & 2/18 \\
(1,0,0) & 0.085633 & 1/18 \\
(1,0,1) & 0.164367 & 2/18 \\
(1,1,0) & 0.164367 & 4/18 \\
(1,1,1) & 0.085633 & 2/18 \\ \bottomrule
\end{tabular}
\end{table}

Both distributions have $\oinfo = -0.072780$ bit, but distinct $\kappa_{\Delta}$ values: Distribution 1 has $\kappa_{\Delta}=3$, while Distribution 2 has $\kappa_{\Delta}\approx2.471$.
Under the standard framing of the O-information as a measure of redundancy/synergy balance, we would say that both distributions have the same balance of higher-order interactions, but the different $\kappa_{\Delta}$ values complicates that interpretation. 
Distribution 1 seems to clearly be the more synergy-dominated one, being composed of a pure, three-way XOR relationship, while Distribution 2 has a mixture of both XOR and COPY dependencies. 
The $\kappa_{\Delta}$ measure distinguishes between these two cases (pure synergy vs a mixture), while O-information does not.

Collectively, these results show that these various features of multivariate distributions (total deviation from independence, redundancy/synergy dominance, behavior under single-element marginalization, etc) are semi-independent aspects of the higher-order structure.
By pinning different aspects while allowing others to vary, we can build a better intuition of what the ``landscape" of higher-order interactions looks like.

\subsubsection{A geometry of higher-order interactions}

\begin{figure}
    \centering
    \includegraphics[width=1\textwidth]{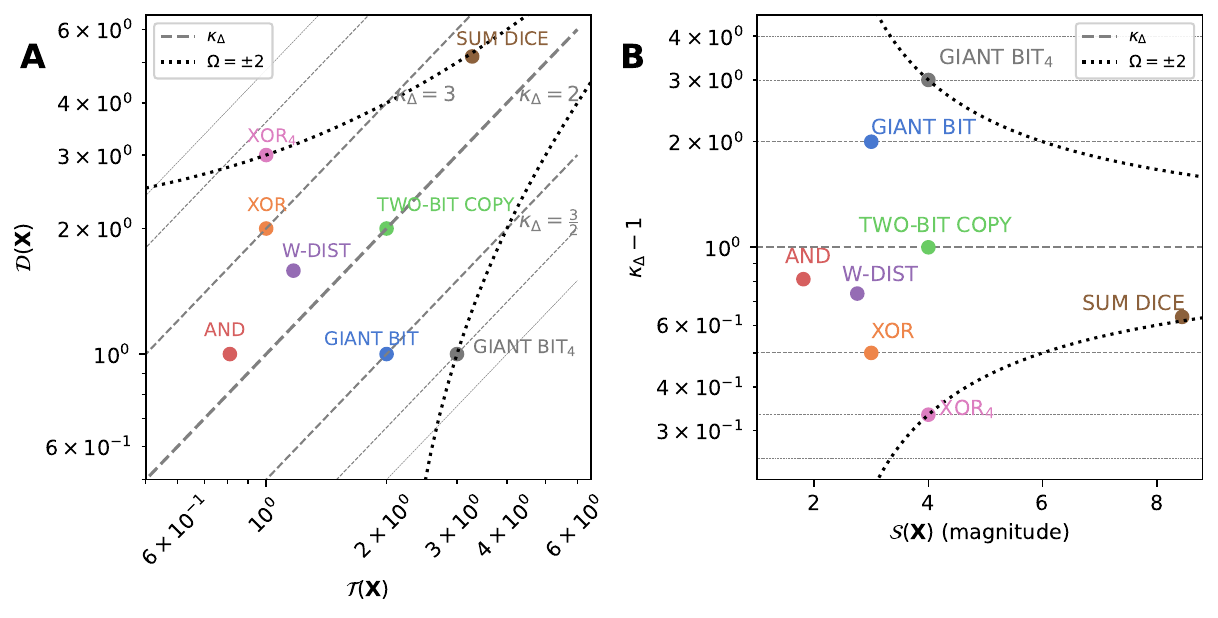}
    \caption{\textbf{A geometry of higher-order interactions. A:} Distributions can be plotted in a $(\tc(\probp),\dtc(\probp))$ plane.
    The dashed rays correspond to fixed values of $\kappa_{\Delta}$, with distance from the origin given by the total dependency (the S-information). 
    The dotted lines correspond to distributions where the O-information is $\pm2$, note that (as in the example given above), multiple distributions can have approximately equal O-informations with distinct values of $\kappa_{\Delta}$ - in this case, XOR$_4$ and the SUMMED\_DICE distributions (XOR$_4$ is a 4-element parity gate, SUMMED\_DICE is a three-dimensional distribution that puts equal probability on all rolls of two fair, six-sided dice).  
    \textbf{B:} Transforming the space into one parameterized by the $\tc/\dtc$ ratio and the S-information creates a space that describes the intensity of dependency on one axis and the redundancy/synergy balance on the other. 
    Again, the dashed lines correspond to fixed ratios, and the dotted lines are $\oinfo(\probp)=\pm2$.}
    \label{fig:delta_k}
\end{figure}

Every distribution $\probp$ can be plotted as a point $(\tc(\xjoint),\dtc(\xjoint))$ in a two-dimensional space, and fixing various features measures corresponds to restricting distributions to a fixed curve (see Fig. \ref{fig:delta_k}A).
Every value of $\kappa_\Delta$ corresponds to a ray emanating from the origin: moving along that ray will change the global intensity of a dependency, but not the ratio that describes the higher-order structure. 
As a result, fixing $\kappa_{\Delta}$ restricts the space of possible distributions to a single ray (the grey, dashed lines).
Likewise, fixing the O-information to a given value restricts the space of possible distributions to a line parallel to the $\kappa_{\Delta}=2$ ray (the black, dotted line). 
Note how XOR$_4$ and SUMMED\_DICE have nearly-equivalent O-informations and fall nearly on the $\oinfo(\probp)=-2$ curve.

\subsection{A conjugate hierarchy of redundancies}
Recently, Rosas et al., introduced the notion of entropic conjugation \cite{rosas_characterising_2025}, an operation that inverts low-order and high-order dependencies.
A natural question is: what is the entropic conjugation of $\Delta^{k}(\probp)$ and how does it behave? 
We will denote $\text{Conj}(\Delta^{k})$ as $\Gamma^{k}(\probp)$ for simplicity's sake. 
Since $\Delta^{k}(\probp) = \sinfo(\probp) - k\tc(\probp)$, it is easy to see that:

\begin{align}
    \Gamma^{k}(\probp) &= \sinfo(\probp)-k\dtc(\probp)
\end{align}

It follows directly that $\Gamma^{k}$ can be arranged into the expected spectrum of $k$ values:

\begin{align}
    \Gamma^{0}(\probp) &= \sinfo(\probp) \\
    \Gamma^{1}(\probp) &= \tc(\probp) \\
    \Gamma^{2}(\probp) &= \oinfo(\probp)
\end{align}

Just as $\Delta^{k}(\probp)$ is arranged into a spectrum of increasingly synergy-weighted measures (by peeling away layers of redundant total correlation from the S-information), $\Gamma^{k}(\probp)$ is arranged into a spectrum of increasingly redundancy-weighted measures (by peeling away layers of synergistic dual total correlation). 
If $\probp$ is composed entirely of redundant dependencies of order $k$ (i.e. a $k$-element giant bit distributions), then $\Gamma^{k}(\probp)=0$ bit (see Proof \ref{proof:gamma_red}). 
This is the mirror image of the behavior of $\Delta^{k}$, but for redundancies of order $k$ rather than synergies of order $k$. 
Since the S-information and the dual total correlation are both additive over independent subsystems, $\Gamma^{k}$ is too, and so $\Gamma^{k}(\probp)=0$ for any system composed of independent giant bits of $k$ elements each.

Straightforward algebra shows that $\Gamma^{k}$ can be written in terms of the total correlation:

\begin{align}
    \Gamma^{k}(\probp) = (1-(N-1)(k-1))\tc(\probp) + (k-1)\sum_{i=1}^{N}\tc(\probp_{\setminus i}) \label{eq:gamma_tc}
\end{align}

This is less than illuminating, however. 
The issue is that it is attempting to describe redundancy in terms of single-element failures. 
This is natural when considering synergies, since single-node failures destroy synergistic dependencies, but it is a poor fit in the case of redundancies, which are, by definition, robust to failures. 

Instead we can ask: what is the operation that ``destroys" redundancy the same way that marginalization ``destroys" synergy?
The obvious answer is: conditioning. 
Algebraic manipulation then reveals the intuitive form:

\begin{equation}
    \Gamma^{k}(\probp)=(N-k)\dtc(\probp)-\sum_{i=1}^{N}\dtc(\probp_{\setminus i\mid i})
\end{equation} 

For a full proof see Proof \ref{proof:gamma_k}. 

In this form, we can see that $\Gamma^{k}$ has the same logical structure as $\Delta^{k}$. If $\probp$ were composed entirely of independent giant bits of $k$ elements each, conditioning on any single element would erase the dual total correlation of the giant bit it belongs to and leave the others intact, so the conditional dual total correlations would sum to $(N-k)\dtc(\probp)$.
The second term is the actual value, and the sign convention follows identically. 
There is also a zero-crossing point analogous to $\kappa_{\Delta}$:

\begin{align}
    \kappa_{\Gamma} &= \frac{\sinfo(\probp)}{\dtc(\probp)} \\
    &= 1 + \frac{\tc(\probp)}{\dtc(\probp)} \\
    &= 1 + \frac{1}{\kappa_{\Delta}-1} = \frac{\kappa_{\Delta}}{\kappa_{\Delta}-1}
\end{align}

Which leads to a nice, unifying identity:

\begin{align}
    (\kappa_{\Delta}-1)(\kappa_{\Gamma}-1)=1
\end{align}

\section{Further generalization}
The $\Delta^{k}$ and $\Gamma^{k}$ functions provide a unifying, general perspective on multivariate mutual information, but they are nevertheless rather weak in-and-of themselves. 
Both boil down to adding factors of $\tc(\probp)$ or $\dtc(\probp)$, and both are fooled by the phenomenon of ``dyadic camouflage" \cite{james_multivariate_2017}, when higher-order dependencies are masked by co-occuring lower-order ones. 
Is it possible to further generalize $\Delta^{k}$ to relax some of these constraints and make it useful in a wider context? 

Formalizing $\Delta^{k}$ in terms of joint and marginal total correlation reveals something interesting about all of the higher-order information measures. 
Recall that $\tc(\probp)=\kl(\probp\| \prod_{i}\probp_i)$, where $\kl$ is the Kullback-Leibler divergence:
\begin{align}
    \kl(\probp\|\probq) = \mathbb{E}_{\probp}\bigg[\log\frac{\probp(\mathbf{x})}{\probq(\mathbf{x})}\bigg]
\end{align}
The total correlation, then, is the information gained when updating from a prior of no interaction between elements to the true joint distribution. 
Built into the total correlation, and all measures derived from it, is an implicitly chosen reference: maximum entropy. 
This is a very strong reference distribution: in many contexts, the natural reference probably isn't a kind of ``ideal gas", but something more constrained. 
What happens if we re-define $\Delta^{k}$ to accept arbitrary Kullback-Leibler divergences, and allowed users to set both the posterior distribution and the reference prior? 
To avoid naming collision with Jansma, who also uses $\Delta$ to define a higher-order measure on the Kullback-Leibler divergence \cite{jansma_mereological_2025}, we will define this relaxation of $\Delta^{k}$ as:
\begin{align}
    \label{eq:lambda_k}
    \Lambda^{k}(\probp\|\probq) &= (N-k)\kl(\probp||\probq) - \sum_{i=1}^{N}\kl(\probp_{\setminus i}||\probq_{\setminus i}) \\
    &= \Lambda^{0}(\probp\|\probq) - k\kl(\probp\|\probq)
\end{align}
The general behavior of $\Lambda^{k}(\probp\|\probq)$ is not immediately obvious, although the specific case of $k=0$ has previously appeared in the literature as the ``erasure divergence", introduced by Verdu and Weissman \cite{verdu2008}. 
The erasure divergence is a non-negative measure (see Proof \ref{proof:lambda0_nonneg}) and 
since $\Lambda^{k}=\Lambda^{0}-k\kl(\probp\|\probq)$, the whole family can be read as the erasure divergence offset by integer multiples of the ordinary divergence.
By analogy with $\kappa_{\Delta}$, the zero crossing of $\Lambda^{k}$ defines an effective order of the update from $\probq$ to $\probp$:
\begin{align}
    \kappa_{\Lambda} = \frac{\Lambda^{0}(\probp\|\probq)}{\kl(\probp\|\probq)}, \nonumber
\end{align}
defined whenever $\probp\neq\probq$. 
When $\probq=\prod_{i}\probp_i$, $\kappa_{\Lambda}=\kappa_{\Delta}$ and $\Lambda^{k}(\probp\|\probq)=\Delta^{k}(\probp)$.

To better understand the general behavior of $\Lambda^{k}$, we can leverage the following decomposition of the $\kl$ to gain some insight:

\begin{equation}
    \label{eq:dkl_decomp}
    \kl(\probp\|\probq) = \bigg[\tc(\probp) - \tc(\probp,\probq)\bigg] + \sum_{i=1}^{N}\kl(\probp_i||\probq_i)
\end{equation}

For proof, see Proof \ref{proof:dkl_decomp}.
Here $\tc(\probp,\probq)$ is the cross total correlation:
\begin{align}
    \tc(\probp,\probq) = \mathbb{E}_{\probp}\bigg[\log\frac{\probq(\mathbf{x})}{\bar{\probq}(\mathbf{x})}\bigg]
\end{align}

Where $\bar{\probq} = \prod_{i}\probq_i$. 
The cross total correlation is the multivariate generalization of the cross mutual information recently introduced by Gohil et al. \cite{gohil2025}, and unlike the total correlation, it is a signed measure. 
It can be negative when $\probp$ places mass on states that violate the dependencies of $\probq$.

Equation \ref{eq:dkl_decomp} shows us that for a multivariate $\xjoint$, we can decompose the information we gain when updating from a prior belief $\probq(\xjoint)$ to posterior belief of $\probp(\xjoint)$ into two parts. 
The first: $\tc(\probp)-\tc(\probp,\probq)$ is the change in our beliefs about the dependency structure of the whole system.
The second term, $\sum_{i}\kl(\probp_i||\probq_i)$ is the ``marginal drift": the change in our beliefs that is attributable to changes in the first-order features of the parts, irrespective of any interaction between them. 
Unfortunately, this decomposition isn't perfect: marginal information can still ``leak" into dependency term via the cross total correlation; for more see Proof \ref{proof:marginal_leak}.

Substituting Eq. \ref{eq:dkl_decomp} into Eq. \ref{eq:lambda_k} finds:

\begin{align}
    \label{eq:final_lambda_k}
    \Lambda^{k}(\probp||\probq)=\bigg[\Delta^{k}(\probp) - \Delta^{k}(\probp,\probq)\bigg] + (1-k)\sum_{i=1}^{N}\kl(\probp_i||\probq_i)
\end{align}

So, for a given measure in the $\Delta^{k}$ family, relaxing the implicit maximum-entropy prior is the difference between that measure, and its cross-entropy based form, plus a marginal drift term scaled by $(1-k)$.

The last term, the marginal drift, presents something of an interpretive difficulty.
One of the major benefits of measures like total correlation, dual total correlation, and O-information is that they are pure measures of dependency between elements and not confounded by the marginal features (this is guaranteed by the implicit prior of constrained maximum entropy). 
Unfortunately, this is not generally the case for $\Lambda^{k}$: as $k$ grows, the marginal drift term becomes increasingly dominant, swamping the difference of $\Delta^{k}$s.
The explicit marginal-drift term vanishes in the special case where $k=1$ (corresponding to the dual total correlation):

\begin{align}
    \label{eq:dtc_pq}
    \dtc(\probp||\probq) &= \Delta^{1}(\probp) - \Delta^{1}(\probp,\probq) \\
    &= \dtc(\probp)-\dtc(\probp,\probq)
\end{align}

{\color{red}
}

Alterately, we could \textit{post-hoc} modify $\Lambda^{k}$ to create a version that is free of the explicit marginal-drift term:

\begin{align}
    \tilde{\Lambda}^{k}(\probp\|\probq)=\Lambda^{k}(\probp\|\probq)-(1-k)\sum_{i=1}^{N}\kl(\probp_i\|\probq_i).
\end{align}

In the case where $k=1$, $\tilde{\Lambda}^{k}(\probp\|\probq)=\Lambda^{k}(\probp\|\probq)$, although this form still inherits the implicit dependence on the marginals of $\probq$ through the cross terms.

Another option might be to put constraints on what prior/posterior combinations are acceptable. 
For example, if we stipulate that the prior and posterior must have the same univariate marginals (as does the total correlation, or higher-order generalizations like the connected information \cite{schneidman_network_2003}, discussed below), then marginal drift term vanishes without needing to be removed. 
Finally, in the case of continuous Gaussian variables, one can avoid the problem entirely by standardizing all variables to zero mean and unit variance, in which case the marginal drift is zero. 
Which, if any, of these options is the most appropriate will be context specific, and depend largely on what the specific research question being asked is, and what kinds of distributions are seen as good models of the underlying data. 

\subsection{Examples}

\begin{table}[h]
\centering
\caption{Distributions with the same $\kl$ but different order-structures.}
\begin{tabular}{@{}cccc@{}}
\toprule
$\xjoint$   & $\probp(\mathbf{x})$ & $\probq^{\mathrm{dyadic}}(\mathbf{x})$ & $\probq^{\mathrm{triadic}}(\mathbf{x})$ \\ \midrule
(0,0,0) & 25/98 & 5/28  & 5/28  \\
(0,0,1) & 5/49  & 5/28  & 1/14  \\
(0,1,0) & 2/49  & 1/14  & 1/14  \\
(0,1,1) & 5/49  & 1/14  & 5/28  \\
(1,0,0) & 2/49  & 1/14  & 1/14  \\
(1,0,1) & 5/49  & 1/14  & 5/28  \\
(1,1,0) & 25/98 & 5/28  & 5/28  \\
(1,1,1) & 5/49  & 5/28  & 1/14   \\ \bottomrule
\end{tabular}
\label{tab:common_posterior}
\caption*{A posterior $\probp(\mathbf{x})$ and two different prior models. 
The first prior, $\probq^{\mathrm{dyadic}}(\mathbf{x})$ is composed of a dyadic interaction on $X_1$ and $X_2$ (by construction). 
The second prior, $\probq^{\mathrm{triadic}}(\mathbf{x})$ has a synergistic parity-gate interaction (by construction). 
All three are constructed such that the Kullback-Leibler divergence from either prior to the shared posterior is the same: only by accounting for higher-order information gain reveals the difference. }
\end{table}


In the $\Delta^{k}$ section, we constructed pairs of distributions that had fixed values for total correlation, O-information, etc but showed different values of $\kappa_{\Delta}$ and had different $\Delta^{k}$ spectra. 
In the context of $\Lambda^{k}$, we need to construct three distributions: one fixed posterior and two different priors, both of which have the same Kullback-Leibler divergence from prior to posterior.
The posterior should include both a mixture of low-order and high-order interactions, while each prior contains either only low-order or only high-order dependencies. 
This way, we can contrast how $\Lambda^{k}$ behaves when an update requires adding low- or higher-order dependencies to a ``model." 
Here, we constructed all three distributions from a common recipe.

The probabilities in the posterior, which contains both dyadic and triadic interactions are given by:

\begin{align}
    \probp(\mathbf{x}) \propto \big(1 + \frac{3}{2}\big[X_1=X_2\big]\big)\big(1+\frac{3}{2}\big[X_1\oplus X_2\oplus X_3 = 0\big]\big)
\end{align}

The first prior, which contains only a dyadic interaction is given by:

\begin{align}
    \probq^{\textnormal{dyadic}}(\mathbf{x}) \propto \big(1 + \frac{3}{2}\big[X_1=X_2\big]\big)
\end{align}

While the second prior, with only a triadic interaction, is given by:

\begin{align}
    \probq^{\textnormal{triadic}}(\mathbf{x}) \propto \big(1+\frac{3}{2}\big[X_1\oplus X_2\oplus X_3 = 0\big]\big)
\end{align}

All distributions were constructed to have the same global Kullback-Leibler divergence: $\kl(\probp\|\probq^{\mathrm{dyadic}}) = \kl(\probp\|\probq^{\mathrm{triadic}}) \approx 0.136879\textrm{ bit}$, but the structure added to the model differs.
When updating from $\probq^{\mathrm{dyadic}}$ to $\probp$, we go from a belief in a dyadic-only distribution to one that mixes dyadic and triadic interactions. 
Conversely, when updating from $\probq^{\mathrm{triadic}}$ to $\probp$, we are updating our beliefs from a purely triadic model to one that mixes different orders. 
For the values of all three distributions, see Table \ref{tab:common_posterior}.

Since the common posterior and two priors do not have the same first-order marginals, we start with the specific case of $k=1$ for this example as explicit marginal-drift term vanishes. 
We find that $\Lambda^{1}(\probp\|\probq^{\mathrm{dyadic}})\approx0.2248$ bit, while $\Lambda^{1}(\probp\|\probq^{\mathrm{triadic}})\approx0.0879$ bit. 
These results reflect the intuition that, while both updates add dependency structure to our model, the update that includes the triadic interaction adds ``more" in a sense, as it is a higher-dimensional constraint, and dual total correlation itself is preferentially sensitive to synergies rather than redundancies \cite{rosas_characterising_2025}.
The distinction between adding low-order and higher-order mechanisms is highlighted by $\Lambda^{2}(\probp\|\probq)$.
From the dyadic prior, the result is positive: $\approx0.0879$ bit, while from the triadic prior, the result is negative: $\approx-0.0489$ bit. 
The sign tracks whether the update includes adding beyond-pairwise structure (in which case, it is positive, just like the negative O-information), or lower-order structure (in which case, it is negative). 
As we will see below, this window in particular mechanisms can help untangle more complex distributions. 

\subsection{\texorpdfstring{$\mathbf{\Lambda^{k}}$ can distinguish dyadic camouflage}{Lambda-k can distinguish dyadic camouflage}}
\label{sec:lambda_camouflage}
A fundamental limitation of almost all existing information-theoretic measures of higher-order dependency is that they are fooled by ``dyadic camouflage" \cite{james_multivariate_2017}.
Dyadic camouflage occurs when an interaction between lower-order and higher-order dependencies creates the illusion of purely pairwise (dyadic) interactions. 
James and Crutchfield demonstrated this using two carefully-constructed distributions: a distribution composed purely of dyadic interactions, and another with a masked triadic interaction, that had identical values on any measure composed of a linear combination of entropies. 
One of the only measures that successfully disambiguated the two distributions was Schneidman's connected information \cite{schneidman_network_2003}, which uses the Kullback-Leibler divergence of the true distribution from a sequence of increasingly constrained maximum entropy distributions to assign dependency to first order, second order, and so on.

{\color{red}
}

\subsubsection{Distinguishing mechanistic components of dyadic camouflage}
Above, the example given in Table \ref{tab:common_posterior}, we showed how $\Lambda^{k}$ can provide insights into what kind of information is gained when updating from a prior to a posterior. 
We can do the same analysis with the triadic distribution from James and Crutchfield. 
The camouflaged triadic distribution was constructed by grouping three pairs of bits into three, four-state macro-variables.
The first set of bits ($x_1,y_1,z_1$) share a three-way, exclusive-or dependency, while the second set of bits ($x_2,y_2,z_2$) share a single redundant giant bit dependency. 
We can ``break" the dyadic camouflage by asking about the nature of the information gained from updating from an all-XOR prior or an all-giant bit prior. 
\begin{enumerate}
    \item[] All-XOR distribution: The first set of three bits are restricted to $x_1\oplus y_1\oplus z_1 = 0$, while the second set of three bits is independent: $x_2\bot y_2\bot z_2$. 
    \item[] All-giant bit (GB) distribution: The second set of three bits are restricted to $x_2=y_2=z_2$, while the first set of three bits is independent: $x_1\bot y_1\bot z_1$. 
\end{enumerate}

Using these two priors, the dyadic camouflage is easily distinguishable: $\Lambda^{2}(\mathrm{Triadic}||\mathrm{All-XOR}) = -1$ bit, while $\Lambda^{2}(\mathrm{Triadic}||\mathrm{All-GB}) = 1$ bit.
Intuitively this makes sense: when updating from a prior belief that the distribution has only third-order synergy (the all-XOR case), we must build in 1 bit of global redundancy, which to $\Lambda^{2}$ is ``seen" as -1 bit. 
Conversely, when updating from a prior belief that the distribution has only redundancy (the all-GB case), we must build in 1 bit of global synergy (the third-order parity dependency), which is ``seen" by $\Lambda^{2}$ as +1 bit. 
So, while the global synergy and redundancy terms interact in a way that makes it impossible to distinguish them when starting from a prior of maximum entropy (implicit in all of the $\Delta^{k}$ and $\Gamma^{k}$ family of measures), this is not the case if one starts from a different set of priors. 

Unfortunately, we could not do the exact same analysis on the dyadic distribution from \cite{james_multivariate_2017}, as the support for that distribution does not match either the All-XOR or All-GB distributions and so the Kullback-Leibler divergence (and by extension, $\Lambda^{k}$) is not well-defined.
We can approximate it, however, by mixing each mechanistic prior with the uniform distribution $\mathcal{U}$, $\probq_{\epsilon} = (1-\epsilon)\probq + \epsilon\,\mathcal{U}$, which gives every state non-zero prior probability, at the cost of adding noise.
The resulting values of $\Lambda^{2}$ are shown in Table \ref{tab:smoothed_priors}.
Across three orders of magnitude of $\epsilon$, the sign pattern is stable and opposite for the two distributions: updating to the triadic distribution is negative from the All-XOR prior and positive from the All-GB prior, while updating to the dyadic distribution shows the reverse.
For the dyadic distribution, the magnitudes grow without bound as $\epsilon\to0$, since $\kl(\probp^{\mathrm{dyadic}}\|\probq_{\epsilon})$ itself diverges, so it is the signs, not the magnitudes, that carry meaning here.
The pairwise reference used above avoids this issue altogether.

\begin{table}[h]
\centering
\caption{$\Lambda^{2}(\probp\|\probq_{\epsilon})$ (bit) for the dyadic and triadic distributions, with each mechanistic prior smoothed by a fraction $\epsilon$ of the uniform distribution.}
\label{tab:smoothed_priors}
\begin{tabular}{@{}lcccc@{}}
\toprule
$\epsilon$ & Dyadic $\|$ All-XOR & Dyadic $\|$ All-GB & Triadic $\|$ All-XOR & Triadic $\|$ All-GB \\ \midrule
0.5   & $+0.208$ & $-0.203$ & $-0.585$ & $+0.433$ \\
0.1   & $+1.198$ & $-1.574$ & $-0.926$ & $+0.890$ \\
0.01  & $+2.826$ & $-3.991$ & $-0.993$ & $+0.989$ \\
0.001 & $+4.483$ & $-6.475$ & $-0.999$ & $+0.999$ \\
0     & $\to+\infty$& $\to-\infty$& $-1$   & $+1$ \\
\bottomrule
\end{tabular}
\end{table}

More generally, it suggests that future work on extracting higher-order dependencies may benefit from taking a more explicitly mechanistic approach of building a lower-order model and assessing how the true system deviates from it. 

\section{Discussion}

In this paper, we began with four, classic measures of multivariate dependence from information theory (total correlation, dual total correlation, O-information, and S-information), and generalized them to explore the underlying, unified structure of multivariate information. 
We showed that three of the four measures (dual total correlation, negative O-information, and S-information) could be written in a common form in terms of the last one: the total correlation, and this general measure, $\Delta^{k}$ describes the rate at which the global dependency structure disintegrates under single-element failures. 
This unifying perspective also has a dual, $\Gamma^{k}$, induced by entropic conjugation \cite{rosas_characterising_2025} which swaps synergies for redundancies while preserving sensitivity to interaction order. 
Finally, by further relaxing the reliance on total correlation as the core of the $\Delta^{k}$ function and replacing it with general Kullback-Leibler divergence, we were able to complete the generalization of $\Delta^{k}$, introducing a new function $\Lambda^{k}$ that quantifies the higher-order structure learned when updating ones beliefs from an arbitrary prior to arbitrary posterior. 
Collectively, these results show that the large family of higher-order information measures are really different ``faces" of an underlying structure.

This approach gives new insights into the notion of ``order" of synergy or redundancy (as quantified by measures like the O-information), and how it diverges from other notions of interaction order (e.g. the order of Schneidman's connected information \cite{schneidman_network_2003}).

The decomposition of the multivariate Kullback-Leibler divergence (Eq. \ref{eq:dkl_decomp}) offers a new way of thinking about how ``learning'' might occur in high-dimensional spaces.
In complex systems, information-theoretic measures of higher-order interaction have focused almost exclusively on single distributions, with an implicit maximum-entropy reference where one is needed.
The exceptions, such as connected information \cite{schneidman_network_2003,amari2001} and geometric integrated information \cite{oizumi2016}, still construct their reference distribution from the system itself.
In contrast, fields concerned with learning have made extensive use of the Kullback-Leibler divergence to describe the relationship between pairs of distributions (e.g. priors and posteriors).
A special case of Eq. \ref{eq:dkl_decomp}, in which the prior is fully factorized and the cross-total correlation vanishes, has been used to encourage disentangled representations in variational autoencoders \cite{chen_isolating_2018}.
Allowing an arbitrary reference, which may carry dependency structure of its own, and extending the decomposition to the full $\Lambda^{k}$ family may make it applicable to a broader set of questions at the intersection of learning theory and complex systems.
A related approach was previously explored in \cite{varley_generalized_2024}, which introduced a generalized partial information decomposition of the Kullback-Leibler divergence into unique, redundant, and synergistic components.
That approach, however, inherits the usual limitation of antichain-lattice-based decompositions: the number of atoms grows super-exponentially with system size, which makes it impractical for all but the smallest systems.
In contrast, $\Lambda^{k}$ requires only $N+1$ divergences (the joint, and each leave-one-out marginal), and so the number of terms grows linearly with system size.
For example, one might ask how artificial or natural learners come to model a complex system: are the first-order marginals learned first, with the global features following, or vice versa? Are higher-order redundancies easier to learn than higher-order synergies?
The $\Lambda^{k}$ family allows us to characterize not just the structure of a complex system, but also how our beliefs about that structure change.

This may be particularly useful when the implicit maximum-entropy reference is not ideal.
Consider neuroscience, where researchers often compare patterns of brain activity across states: normal consciousness versus anesthesia \cite{luppi2022,varley_consciousness_2020}, psychedelics \cite{peill_psychedelics_2024,timmermann2023}, or healthy versus disease states \cite{herzog_genuine_2022}.
Comparisons built on information-theoretic measures of pairwise or higher-order functional connectivity perform an implicit difference of differences: ``how does Condition X differ from maximum entropy, how does Condition Y differ from maximum entropy, and how do those differences differ?''
The maximum-entropy reference is common to both conditions, but it is not a natural one: the brain is not, and never could be, an ideal gas with no inter-regional interactions.
Worse, the double contrast can be blind to genuine reorganization: as the dyadic and triadic distributions show, two systems can have identical values of every entropy-based measure, yet be distinguished by $\Lambda^{k}$ with an appropriate reference (Section \ref{sec:lambda_camouflage}).
The $\Lambda^{k}$ measure makes a direct comparison possible: we can ask how our model of the brain changes if our prior is State X and our posterior is State Y, $\Lambda^{k}(\probp_{Y}\|\probp_{X})$.
For example, rather than asking ``how do normal consciousness and the psychedelic state each differ from maximum entropy?'', we might ask ``how does my model of the brain change if I initially believe it is in a normal waking state, and then learn that it is under the influence of LSD?''

One interesting feature of this formalism is that the mechanism of $\Delta^{k}(\xjoint)$ is purely combinatorial: counting how many times an interaction of order $k$ survives removing single elements. 
The fact that the measure is total correlation turns out to be somewhat arbitrary: in theory, any function $f(\xjoint)$ could be used, as long as three criteria are met:
\begin{enumerate}
    \item Non-negativity: $f(\xjoint) \geq 0$.
    \item Monotonicity under marginalization: $f(\xjoint) \geq f(\xjoint^{-i})$.
    \item Additivity over independent parts: if $\xjoint\perp\yjoint$, then $f(\xjoint,\yjoint)=f(\xjoint)+f(\yjoint)$.
\end{enumerate}
The notion of a pure interaction is then defined by $f$ rather than assumed: $\xjoint$ is an $f$-pure interaction of order $|\xjoint|$ if $f(\xjoint)>0$ and $f(\xjoint^{-i})=0$ for every $i$.
Under these conditions, the arguments of Proofs \ref{proof:additive_delta_k} and \ref{proof:pure_subsystems} go through unchanged, and $\Delta^{k}_{f}$ vanishes on any system composed of independent $f$-pure interactions of order $k$.

Given those desiderata, the choice of function $f$ merely tunes the ``kind" of dependency that $\Delta^{k}$ is sensitive to.
This allows us to extend this approach to thinking about synergistic interactions in multivariate systems beyond information theory specifically: many branches of mathematics have set functions that satisfy these definitions.
One easily accessible example is the graph cyclomatic number (also known as the first Betti number of the graph). 
The graph cyclomatic number defines a cycle as a ``pure" synergistic relationship in a graph, and measures analogous to dual total correlation, S-information, and O-information can be constructed from it via an analog of $\Delta^{k}$.
See Appendix 2 for further discussion. 
One natural family of such functions is the polymatroid nullities: Shannon entropy is a polymatroid \cite{FUJISHIGE197855}, the total correlation is its nullity, and the graph cyclomatic number is the nullity of a graphic matroid \cite{oxley2011}. 
We leave a general treatment of $\Delta^{k}$ and $\Gamma^{k}$ in this setting to future work.

\section{Conclusions}

The S-information, the dual total correlation, and the negative O-information are usually treated as distinct measures, each with its own motivation and interpretation.
Here we have shown that they are special case of a common general form, $\Delta^{k}$, which compares the total correlation of the whole system to the total correlations of its leave-one-out marginals.
The parameter $k$ sets an expected degree of fragility under single-element failure, and the point at which $\Delta^{k}$ crosses zero, $\kappa_{\Delta}$, defines an effective order of the system's dependencies.
Entropic conjugation yields a twin family, $\Gamma^{k}$, which plays the same role for redundancies, with conditioning taking the place of marginalization.
We further generalized $\Delta^{k}$, introducing $\Lambda^{k}(\probp\|\probq)$, which applies the same leave-one-out construction to the Kullback-Leibler divergence between an arbitrary prior and posterior.
With the correct priors, $\Lambda^{k}$ can distinguish between dyadically camouflaged distributions, and identify the missing mechanistic components.
More generally, $\Lambda^{k}$ shifts the object of study from the structure of a single distribution to the structure of an update: what kind of higher-order information is gained when updating ones beliefs from one model to another?
We hope that treating these measures as different faces of a single structure will make them easier to interpret, compare, and extend.

\subsection*{Acknowledgments}
This work received no special funding. 
The author would like to acknowledge Josh Bonagrd and the Morphology, Evolution, and Cognition Lab at UVM for feedback on some of these ideas. 
The author would also like to acknowledge Olaf Sporns for discussion of the graph cyclomatic number results in particular. 

\subsection*{Author contributions}
All work was done by TFV. 

\subsection*{Code availability}
Code to reproduce all of the examples in this paper is available as a supplementary Python script: \texttt{examples\_final.py}. 

\subsection*{Statement of AI use}
The Claude family of models were used as proof assistants, and for constructing targeted example distributions.
All figures were made by the author, all statements and claims have been verified by the author manually. 

\bibliography{many_faces}

\newpage 

\appendix
\section{Proofs}

  \subsection{Writing DTC in terms of TC}
  \label{proof:dtc_from_tc}

  \begin{theorem}
  The dual total correlation can be expressed as a linear combination of joint and marginal total correlations:
  \begin{align*}
      \dtc(\probp) = (N-1)\tc(\probp) - \sum_{i=1}^{N}\tc(\probp_{\setminus i})
  \end{align*}
  \end{theorem}

  \begin{proof}
  We expand the right-hand side using the definition $\tc(\xjoint) = \sum_{i=1}^{N}\ent(\probp_i) - \ent(\probp)$:
  \begin{align*}
      (N-1)\tc(\probp) - \sum_{i=1}^{N}\tc(\probp_{\setminus i})
      &= (N-1)\left[\sum_{i=1}^{N}\ent(\probp_i) - \ent(\probp)\right] - \sum_{i=1}^{N}\left[\sum_{j \neq
  i}\ent(\probp_j) - \ent(\probp_{\setminus i})\right]
  \end{align*}

  Expanding and regrouping:
  \begin{align*}
      &= (N-1)\sum_{i=1}^{N}\ent(\probp_i) - (N-1)\ent(\probp) - \sum_{i=1}^{N}\sum_{j \neq i}\ent(\probp_j) +
  \sum_{i=1}^{N}\ent(\probp_{\setminus i})
  \end{align*}

  The double sum $\sum_{i=1}^{N}\sum_{j \neq i}\ent(\probp_j)$ counts each $\ent(\probp_j)$ exactly $N-1$ times,
  yielding:
  \begin{align*}
      &= (N-1)\sum_{i=1}^{N}\ent(\probp_i) - (N-1)\ent(\probp) - (N-1)\sum_{i=1}^{N}\ent(\probp_i) +
  \sum_{i=1}^{N}\ent(\probp_{\setminus i}) \\
      &= \sum_{i=1}^{N}\ent(\probp_{\setminus i}) - (N-1)\ent(\probp)
  \end{align*}

  This is precisely Han's definition of the dual total correlation \cite{han_nonnegative_1978}.
  \end{proof}

  \subsection{\texorpdfstring{$\mathbf{\Delta^{k}}$ is additive over independent subsets.}{Delta-k is additive over independent subsets.}}
  \label{proof:additive_delta_k}

  \begin{theorem}
  The function $\Delta^k$ is additive over independent subsystems: if $\xjoint \perp \yjoint$, then
  \begin{align*}
      \Delta^k(\xjoint, \yjoint) = \Delta^k(\xjoint) + \Delta^k(\yjoint)
  \end{align*}
  \end{theorem}

  \begin{proof}
  Let $|\xjoint| = M$ and $|\yjoint| = N$. Since $\xjoint \perp \yjoint$, we have $\tc(\xjoint, \yjoint) =
  \tc(\xjoint) + \tc(\yjoint)$.

  Applying the definition of $\Delta^k$ to the joint system:
  \begin{align*}
      \Delta^k(\xjoint, \yjoint) = (M + N - k)\tc(\xjoint, \yjoint) - \sum_{i=1}^{M+N}\tc((\xjoint, \yjoint)^{-i})
  \end{align*}

  The sum over leave-one-out marginals partitions into removals from $\xjoint$ and removals from $\yjoint$. When
  element $i$ is removed from $\xjoint$, the marginal total correlation is $\tc(\xjoint^{-i}) + \tc(\yjoint)$;
  similarly for removals from $\yjoint$. Thus:
  \begin{align*}
      \sum_{i=1}^{M+N}\tc((\xjoint, \yjoint)^{-i})
      &= \sum_{i=1}^{M}\left[\tc(\xjoint^{-i}) + \tc(\yjoint)\right] + \sum_{j=1}^{N}\left[\tc(\xjoint) +
  \tc(\yjoint^{-j})\right] \\
      &= \sum_{i=1}^{M}\tc(\xjoint^{-i}) + M\tc(\yjoint) + N\tc(\xjoint) + \sum_{j=1}^{N}\tc(\yjoint^{-j})
  \end{align*}

  Substituting and expanding $(M + N - k)[\tc(\xjoint) + \tc(\yjoint)]$:
  \begin{align*}
      \Delta^k(\xjoint, \yjoint)
      &= (M + N - k)\tc(\xjoint) + (M + N - k)\tc(\yjoint) \\
      &\quad - \sum_{i=1}^{M}\tc(\xjoint^{-i}) - M\tc(\yjoint) - N\tc(\xjoint) - \sum_{j=1}^{N}\tc(\yjoint^{-j})
  \end{align*}

  Collecting terms:
  \begin{align*}
      &= (M - k)\tc(\xjoint) - \sum_{i=1}^{M}\tc(\xjoint^{-i}) + (N - k)\tc(\yjoint) -
  \sum_{j=1}^{N}\tc(\yjoint^{-j}) \\
      &= \Delta^k(\xjoint) + \Delta^k(\yjoint)
  \end{align*}
  \end{proof}

  \subsection{Purely-synergistic subsystems}
  \label{proof:pure_subsystems}

  \begin{theorem}
  If $\xjoint$ is composed entirely of synergistic interactions of order $k$, then $\Delta^k(\xjoint) = 0$.
  \end{theorem}

  \begin{proof}
  Let $\xjoint$ be partitioned into $m$ mutually independent subsets $\{\xjoint_{\alpha_1}, \xjoint_{\alpha_2},
  \ldots, \xjoint_{\alpha_m}\}$, where each $\xjoint_{\alpha_i}$ instantiates a synergistic interaction of order
  $k$. By Proof \ref{proof:additive_delta_k}, $\Delta^k$ is additive over independent subsystems:
  \begin{align*}
      \Delta^k(\xjoint) = \sum_{i=1}^{m} \Delta^k(\xjoint_{\alpha_i})
  \end{align*}

  It therefore suffices to show that $\Delta^k(\xjoint_{\alpha_i}) = 0$ for a single $k$-element synergistic
  interaction.

  Since $|\xjoint_{\alpha_i}| = k$:
  \begin{align*}
      \Delta^k(\xjoint_{\alpha_i}) = (k - k)\tc(\xjoint_{\alpha_i}) - \sum_{j=1}^{k}\tc(\xjoint_{\alpha_i}^{-j}) =
  -\sum_{j=1}^{k}\tc(\xjoint_{\alpha_i}^{-j})
  \end{align*}
  By definition, a synergistic interaction of order $k$ satisfies $\tc(\xjoint_{\alpha_i}^{-j}) = 0$ for all $j$:
  removing any single element destroys the dependency entirely. Therefore:
  \begin{align*}
      \Delta^k(\xjoint_{\alpha_i}) = 0
  \end{align*}
  \end{proof}

  \subsection{Purely-redundant subsystems}
  \label{proof:gamma_red}

  \begin{theorem}
  If $\xjoint$ is composed entirely of redundant interactions of order $k$, then $\Gamma^k(\xjoint) = 0$.
  \end{theorem}

  \begin{proof}
  As with $\Delta^k$, the function $\Gamma^k$ is additive over independent subsystems. It therefore suffices to
  prove the result for a single $k$-element redundant interaction: a system of $k$ identical copies of a single
  random variable (a ``giant bit'').

  For such a system, we have the following identities:
  \begin{align*}
      \tc(\xjoint) &= (k-1)\ent(X_1) \\
      \dtc(\xjoint) &= \ent(X_1) \\
      \sinfo(\xjoint) &= \tc(\xjoint) + \dtc(\xjoint) = k\ent(X_1)
  \end{align*}

  Applying the definition $\Gamma^k(\xjoint) = \sinfo(\xjoint) - k\dtc(\xjoint)$:
  \begin{align*}
      \Gamma^k(\xjoint) = k\ent(X_1) - k\ent(X_1) = 0
  \end{align*}
  \end{proof}

  \subsection{\texorpdfstring{$\mathbf{\Gamma^{k}}$ can be written in terms of conditional dual total correlation.}{Gamma-k can be written in terms of conditional dual total correlation.}}
  \label{proof:gamma_k}

  \begin{theorem}
  $\Gamma^{k}(\xjoint) = (N-k)\dtc(\xjoint) - \sum_{i=1}^{N}\dtc(\xjoint^{-i} | X_i)$
  \end{theorem}

  \begin{proof}
  The proof follows from straightforward information-theoretic identities:
  \begin{align*}
      \textrm{Let: } \Gamma^{k}(\xjoint) &= (N-k)\dtc(\xjoint) - \sum_{i=1}^{N}\dtc(\xjoint^{-i} | X_i)
  \end{align*}

  We begin by expanding the conditional dual total correlation term into entropies:
  \begin{align*}
      &= (N-k)\dtc(\xjoint) - \sum_{i=1}^{N}\bigg[\ent(\xjoint^{-i}|X_i) -
  \sum_{j\not=i}\ent(X_j|\xjoint^{-ij},X_i)\bigg]
  \end{align*}

  We can collapse $\xjoint^{-ij},X_i$ into $\xjoint^{-j}$:
  \begin{align*}
      &= (N-k)\dtc(\xjoint) - \sum_{i=1}^{N}\bigg[\ent(\xjoint) - \ent(X_i) -
  \sum_{j\not=i}\ent(X_j|\xjoint^{-j})\bigg] \\
      &= (N-k)\dtc(\xjoint) - \bigg[\sum_{i=1}^{N}\ent(\xjoint) - \sum_{i=1}^{N}\ent(X_i) -
  \sum_{i=1}^{N}\sum_{j\not=i}\ent(X_j|\xjoint^{-j})\bigg]
  \end{align*}

  The value of the double-summation depends only on $X_j$ and so each conditional entropy \strike{appear} {\color{red}appears} $N-1$ times:
  \begin{align*}
      &= (N-k)\dtc(\xjoint) - \bigg[N\ent(\xjoint) - \sum_{i=1}^{N}\ent(X_i) -
  (N-1)\sum_{i=1}^{N}\ent(X_i|\xjoint^{-i})\bigg] \\
      &= (N-k)\dtc(\xjoint) - N\ent(\xjoint) + \sum_{i=1}^{N}\ent(X_i) + (N-1)\sum_{i=1}^{N}\ent(X_i|\xjoint^{-i})
  \end{align*}

  As an interesting consequence, the term $\sum_{i=1}^{N}\dtc(\xjoint^{-i}|X_i) = (N-1)\dtc(\xjoint) -
  \tc(\xjoint)$.

  We now expand the first dual total correlation term:
  \begin{align*}
      &= (N-k)\ent(\xjoint) - (N-k)\sum_{i=1}^{N}\ent(X_i|\xjoint^{-i}) - N\ent(\xjoint) + \sum_{i=1}^{N}\ent(X_i)
  + (N-1)\sum_{i=1}^{N}\ent(X_i|\xjoint^{-i})
  \end{align*}

  Joint entropy terms cancel:
  \begin{align*}
      &= -k\ent(\xjoint) - (N-k)\sum_{i=1}^{N}\ent(X_i|\xjoint^{-i}) + \sum_{i=1}^{N}\ent(X_i) +
  (N-1)\sum_{i=1}^{N}\ent(X_i|\xjoint^{-i}) \\
      &= -k\ent(\xjoint) + \sum_{i=1}^{N}\ent(X_i) + (k-N)\sum_{i=1}^{N}\ent(X_i|\xjoint^{-i}) +
  (N-1)\sum_{i=1}^{N}\ent(X_i|\xjoint^{-i}) \\
      &= -k\ent(\xjoint) + \sum_{i=1}^{N}\ent(X_i) + (k-1)\sum_{i=1}^{N}\ent(X_i|\xjoint^{-i}) \\
      &= -k\ent(\xjoint) + \sum_{i=1}^{N}\ent(X_i) + k\sum_{i=1}^{N}\ent(X_i|\xjoint^{-i}) -
  \sum_{i=1}^{N}\ent(X_i|\xjoint^{-i}) \\
      &= \sum_{i=1}^{N}\ent(X_i) - \sum_{i=1}^{N}\ent(X_i|\xjoint^{-i}) - k\ent(\xjoint) +
  k\sum_{i=1}^{N}\ent(X_i|\xjoint^{-i})
  \end{align*}

  From the identity that $\dtc(\xjoint) = \ent(\xjoint) - \sum_{i=1}^{N}\ent(X_i|\xjoint^{-i})$ we can simplify:
  \begin{align*}
      &= \sum_{i=1}^{N}\mi(X_i;\xjoint^{-i}) - k\dtc(\xjoint)
  \end{align*}

  By the identity that $\sinfo(\xjoint) = \sum_{i=1}^{N}\mi(X_i;\xjoint^{-i})$ we finally show:
  \begin{align*}
      &= \sinfo(\xjoint) - k\dtc(\xjoint) \tag*{\qedhere}
  \end{align*}
  \end{proof}

  \subsection{Decomposition of multivariate Kullback-Leibler divergence.}
  \label{proof:dkl_decomp}

  \begin{theorem}
  $\kl(\probp\|\probq) = \bigg[\tc(\probp) - \tc(\probp, \probq)\bigg] + \sum_{i=1}^{N}\kl(\probp_i\|\probq_i)$
  \end{theorem}

  \begin{proof}
  Let $\bar{\probp}(\xjoint) = \prod_{i=1}^{N}\probp_i(x_i)$ and likewise $\bar{\probq}(\xjoint) = \prod_{i=1}^{N}\probq_i(x_i)$.

  Recall the definition of the Kullback-Leibler divergence:
  \begin{align*}
      \kl(\probp\|\probq) = \mathbb{E}_P\bigg[\log\frac{\probp(\mathbf{x})}{\probq(\mathbf{x})}\bigg]
  \end{align*}

  Expand the probability ratios:
  
  \begin{align*}
      \frac{\probp(\mathbf{x})}{\probq(\mathbf{x})} = \frac{\probp(\mathbf{x})}{\bar{\probp}(\mathbf{x})} \times
  \frac{\bar{\probp}(\mathbf{x})}{\bar{\probq}(\mathbf{x})} \times \frac{\bar{\probq}(\mathbf{x})}{\probq(\mathbf{x})}
  \end{align*}

  The logarithm turns multiplication into addition:
  \begin{align*}
      \log\frac{\probp(\mathbf{x})}{\probq(\mathbf{x})} = \log\frac{\probp(\mathbf{x})}{\bar{\probp}(\mathbf{x})} +
  \log\frac{\bar{\probp}(\mathbf{x})}{\bar{\probq}(\mathbf{x})} + \log\frac{\bar{\probq}(\mathbf{x})}{\probq(\mathbf{x})}
  \end{align*}

  By linearity of expectation:
  \begin{align*}
      \mathbb{E}_{P}\bigg[\log\frac{\probp(\mathbf{x})}{\probq(\mathbf{x})}\bigg] =
  \mathbb{E}_{P}\bigg[\log\frac{\probp(\mathbf{x})}{\bar{\probp}(\mathbf{x})}\bigg] +
  \mathbb{E}_{P}\bigg[\log\frac{\bar{\probp}(\mathbf{x})}{\bar{\probq}(\mathbf{x})}\bigg] +
  \mathbb{E}_{P}\bigg[\log\frac{\bar{\probq}(\mathbf{x})}{\probq(\mathbf{x})}\bigg]
  \end{align*}

  The first term on the RHS is the total correlation of $\xjoint$ computed with respect to $P$:
  \begin{align*}
      \tc(\probp) = \mathbb{E}_P\bigg[\log\frac{\probp(\mathbf{x})}{\bar{\probp}(\mathbf{x})}\bigg]
  \end{align*}

  The last term on the RHS is the negative cross-total correlation: the expected value with respect to $\probp$ of
  the local total correlation of $\probq$:
  \begin{align*}
      \tc(\probp, \probq) = \mathbb{E}_{P}\bigg[\log\frac{\probq(\mathbf{x})}{\bar{\probq}(\mathbf{x})}\bigg]
  \end{align*}

  So we have a decomposition of the Kullback-Leibler divergence into:
  \begin{align*}
      \kl(P || Q) = \bigg[\tc(\probp) - \tc(\probp, \probq)\bigg] +
  \mathbb{E}_{P}\bigg[\log\frac{\bar{\probp}(\mathbf{x})}{\bar{\probq}(\mathbf{x})}\bigg]
  \end{align*}

  To simplify the final term, recognize that $\bar{\probp}$ and $\bar{\probq}$ both factorize:
  \begin{align*}
      \log\frac{\bar{\probp}(\mathbf{x})}{\bar{\probq}(\mathbf{x})} = \sum_{i=1}^{N}\log\frac{\probp_i(x_i)}{\probq_i(x_i)}
  \end{align*}

  Wrap this in an expected value and we get:
  \begin{align*}
      \mathbb{E}_P\bigg[\log\frac{\bar{\probp}(\mathbf{x})}{\bar{\probq}(\mathbf{x})}\bigg] &=
  \sum_{i=1}^{N}\mathbb{E}_{P_{i}}\bigg[\log\frac{\probp_i(x_i)}{\probq_i(x_i)}\bigg] \\
      &= \sum_{i=1}^{N}\kl(\probp_i \| \probq_i)
  \end{align*}

  This term is the ``marginal drift'': the sum of divergences of all the individual $X_i$. 
  It is the amount of information gain attributable to changing your beliefs about the individual elements alone,
  independent of any interactions they may participate in.

  \textbf{Note:} If your prior $Q$ and posterior $P$ have the same first-order marginals, then this final term
  vanishes.

  So, our final decomposition is:
  \begin{align*}
      \kl(\probp \| \probq) = \bigg[\tc(\probp) - \tc(\probp, \probq)\bigg] + \sum_{i=1}^{N}\kl(\probp_i \| \probq_i)
  \end{align*}
  \end{proof}

\subsection{\texorpdfstring{$\mathbf{\Lambda^{0}}$ is non-negative.}{Lambda-0 is non-negative.}}
\label{proof:lambda0_nonneg}

The quantity $\Lambda^{0}$ coincides with the erasure divergence of Verdu and Weissman \cite{verdu2008}.

\begin{theorem}
For any $\probp$ and $\probq$ over $\xjoint$ with $\kl(\probp\|\probq) < \infty$, $\Lambda^{0}$ is a sum of expected conditional Kullback-Leibler divergences, and is therefore non-negative:
\begin{align*}
    \Lambda^{0}(\probp\|\probq) = \sum_{i=1}^{N}\mathbb{E}_{\probp^{-i}}\bigg[\kl\Big(\probp(X_i|\xjoint^{-i})\,\Big\|\,\probq(X_i|\xjoint^{-i})\Big)\bigg] \geq 0
\end{align*}
Moreover, $\Lambda^{0}(P\|Q) = 0$ if and only if, for every $i$, $\probp(X_i|\xjoint^{-i}) = \probq(X_i|\xjoint^{-i})$ for $\probp^{-i}$-almost every $\xjoint^{-i}$.
\end{theorem}

\begin{proof}
Recall the definition of $\Lambda^{k}$ at $k=0$:
\begin{align*}
    \Lambda^{0}(P\|Q) = N\,\kl(P\|Q) - \sum_{i=1}^{N}\kl(\probp^{-i}\|\probq^{-i})
\end{align*}

Since $N\,\kl(P\|Q) = \sum_{i=1}^{N}\kl(P\|Q)$, we can pair each leave-one-out term with one copy of the joint divergence:
\begin{align*}
    \Lambda^{0}(P\|Q) = \sum_{i=1}^{N}\bigg[\kl(P\|Q) - \kl(\probp^{-i}\|\probq^{-i})\bigg]
\end{align*}

It therefore suffices to show that each bracketed term is a non-negative expectation of conditional divergences. Fix $i$. We assume $\kl(P\|Q)<\infty$, so $\probp(\mathbf{x})>0$ implies $\probq(\mathbf{x})>0$, and likewise for the marginals $\probp^{-i}$ and $\probq^{-i}$. Both conditionals are therefore well-defined wherever $P$ has support.

Factorize each distribution into a marginal and a conditional:
\begin{align*}
    \probp(\mathbf{x}) = \probp^{-i}(\mathbf{x}^{-i})\,\probp(x_i|\mathbf{x}^{-i}), \qquad \probq(\mathbf{x}) = \probq^{-i}(\mathbf{x}^{-i})\,\probq(x_i|\mathbf{x}^{-i})
\end{align*}

The logarithm turns the ratio of products into a sum:
\begin{align*}
    \log\frac{\probp(\mathbf{x})}{\probq(\mathbf{x})} = \log\frac{\probp^{-i}(\mathbf{x}^{-i})}{\probq^{-i}(\mathbf{x}^{-i})} + \log\frac{\probp(x_i|\mathbf{x}^{-i})}{\probq(x_i|\mathbf{x}^{-i})}
\end{align*}

By linearity of expectation:
\begin{align*}
    \kl(P\|Q) = \mathbb{E}_{P}\bigg[\log\frac{\probp^{-i}(\mathbf{x}^{-i})}{\probq^{-i}(\mathbf{x}^{-i})}\bigg] + \mathbb{E}_{P}\bigg[\log\frac{\probp(x_i|\mathbf{x}^{-i})}{\probq(x_i|\mathbf{x}^{-i})}\bigg]
\end{align*}

The first term depends on $\mathbf{x}$ only through $\mathbf{x}^{-i}$, so its expectation under $P$ is its expectation under $\probp^{-i}$. This is exactly $\kl(\probp^{-i}\|\probq^{-i})$.

For the second term, condition on $\mathbf{x}^{-i}$ using the law of total expectation:
\begin{align*}
    \mathbb{E}_{P}\bigg[\log\frac{\probp(x_i|\mathbf{x}^{-i})}{\probq(x_i|\mathbf{x}^{-i})}\bigg]
    = \mathbb{E}_{\probp^{-i}}\Bigg[\sum_{x_i}\probp(x_i|\mathbf{x}^{-i})\log\frac{\probp(x_i|\mathbf{x}^{-i})}{\probq(x_i|\mathbf{x}^{-i})}\Bigg]
    = \mathbb{E}_{\probp^{-i}}\bigg[\kl\Big(\probp(X_i|\xjoint^{-i})\,\Big\|\,\probq(X_i|\xjoint^{-i})\Big)\bigg]
\end{align*}

Rearranging gives the chain rule for the Kullback-Leibler divergence:
\begin{align*}
    \kl(P\|Q) - \kl(\probp^{-i}\|\probq^{-i}) = \mathbb{E}_{\probp^{-i}}\bigg[\kl\Big(\probp(X_i|\xjoint^{-i})\,\Big\|\,\probq(X_i|\xjoint^{-i})\Big)\bigg]
\end{align*}

The quantity inside the expectation is a Kullback-Leibler divergence between two distributions over $X_i$, so it is non-negative by Gibbs' inequality. Its expectation is therefore non-negative. Substituting back into the sum over $i$:
\begin{align*}
    \Lambda^{0}(P\|Q) = \sum_{i=1}^{N}\mathbb{E}_{\probp^{-i}}\bigg[\kl\Big(\probp(X_i|\xjoint^{-i})\,\Big\|\,\probq(X_i|\xjoint^{-i})\Big)\bigg] \geq 0
\end{align*}

A sum of non-negative terms vanishes if and only if every term does. Each term vanishes if and only if the inner divergence is zero for $\probp^{-i}$-almost every $\mathbf{x}^{-i}$, which holds if and only if the two conditionals are equal there. \qedhere
\end{proof}

\textbf{Note:} $\Lambda^{0} = 0$ does not imply $P = Q$. Let $P$ and $Q$ both be supported on $\{000, 111\}$ with different weights. Every full conditional is deterministic and identical under the two distributions, so $\Lambda^{0}=0$ even though $\kl(P\|Q)>0$. Matching the full conditionals does not fix the weights on the connected components of the support.

\textbf{Note:} The result is specific to $k=0$. $\Lambda^{1}$ and $\Lambda^{2}$ etc can be negative.

\subsection{Marginal changes can leak into the cross total-correlation term}
\label{proof:marginal_leak}

Let $\probp$ and $\probq$ be $N$-element giant bits with $\probp(1,\ldots,1)=p$, $\probp(0,\ldots,0)=1-p$, $\probq(0,\ldots,0)=1-q$ and $\probq(1,\ldots,1)=q$. 
The Kullback-Leibler divergence $\kl(\probp\|\probq)=p\log\frac{p}{q}+(1-p)\log\frac{1-p}{1-q}$.

Every univariate marginal is a binary variable with the same bias as the giant bit, and every leave-one-out marginal is an $(N-1)$-element giant bit with that same bias again. 
As a result, each of these marginals has the same Kullback-Leibler divergence.
Hence, by Eqs. \ref{eq:dkl_decomp} and \ref{eq:lambda_k},
\begin{align*}
    &\sum_{i=1}^{N}\kl(\probp_i\|\probq_i) = N\times\kl(\probp\|\probq), \\ 
    &\tc(\probp)-\tc(\probp,\probq) = -(N-1)\times\kl(\probp\|\probq), \\ 
    &\Lambda^{k}(\probp\|\probq) = -k\times\kl(\probp\|\probq).
\end{align*}
The dependency structure is identical, yet the dependency term is non-zero: the single change in the shared bias is counted once per marginal, and the dependency term subtracts the excess.

\newpage

\section{Graph cyclomatic number as a case study}

Previously, we have speculated about adapting mathematical frameworks from information theory to graph invariants, and analyzed one candidate \cite{varley_scalable_2024}.
Here we propose another: the cyclomatic number.
For a loopless, undirected graph (with potentially parallel edges) $G = (V, E)$, the cyclomatic number is:
\begin{equation}
    \mathcal{C}(G) = |E| - |V| + c(G) \label{eq:cyclomatic_num}
\end{equation}
where $|E|$ is the number of edges, $|V|$ is the number of vertices, and $c(G)$ is the number of connected components.
The cyclomatic number is also the first Betti number of $G$.
We write $G^{-e}$ for the graph with edge $e$ deleted; all vertices are retained.
An edge is a \textit{bridge} if it lies on no cycle (equivalently, if deleting it increases the number of connected components), and we write $b(G)$ for the number of bridges in $G$.

\subsubsection*{The cyclomatic number satisfies the desiderata}

Non-negativity is immediate: $\mathcal{C}(G)$ is the dimension of a vector space.
For monotonicity, consider deleting a single edge $e$.
If $e$ lies on a cycle, then $|E|$ decreases by one, $c(G)$ is unchanged, and one independent cycle is destroyed.
If $e$ is a bridge, then $|E|$ decreases by one and $c(G)$ increases by one, so the cyclomatic number is unchanged:
\begin{align}
    \mathcal{C}(G^{-e}) =
    \begin{cases}
        \mathcal{C}(G) - 1 & \text{if $e$ lies on a cycle,} \\
        \mathcal{C}(G) & \text{if $e$ is a bridge.}
    \end{cases}
    \label{eq:edge_deletion}
\end{align}
In either case $\mathcal{C}(G^{-e}) \leq \mathcal{C}(G)$.
The cyclomatic number is also additive over disjoint subgraphs, since each of $|E|$, $|V|$, and $c(G)$ is.

Finally, $G$ is a pure interaction in the sense of the main text if $\mathcal{C}(G) > 0$ and $\mathcal{C}(G^{-e}) = 0$ for every edge $e$.
By Eq. \ref{eq:edge_deletion}, this requires every edge to lie on every cycle, which holds if and only if the edge set of $G$ forms a single cycle.
The pure interactions of order $k$ are therefore exactly the cycles of $k$ edges: the circuits of the graph matroid.
A pair of parallel edges is a cycle of length two, and is the graphical analogue of a two-element ``giant bit''.

\subsubsection*{The general form}

Replacing total correlation with the cyclomatic number, and elements with edges, gives:
\begin{align}
    \Delta^{k}_{\beta}(G) = (|E| - k)\,\mathcal{C}(G) - \sum_{e \in E}\mathcal{C}(G^{-e})
\end{align}
By Eq. \ref{eq:edge_deletion}, the leave-one-out sum is $|E|\,\mathcal{C}(G) - \big(|E| - b(G)\big)$, since each of the $|E| - b(G)$ non-bridge edges reduces the cyclomatic number by exactly one.
Substituting:
\begin{align}
    \Delta^{k}_{\beta}(G) = \big(|E| - b(G)\big) - k\,\mathcal{C}(G) \label{eq:delta_beta_closed}
\end{align}
Like the information-theoretic $\Delta^{k}$, this is an affine function of $k$, with intercept $\sinfo_{\beta}(G)$ and slope $-\mathcal{C}(G)$.
It crosses zero at:
\begin{align}
    \kappa^{\star}_{\beta}(G) = \frac{|E| - b(G)}{\mathcal{C}(G)}
\end{align}
which is the number of edges lying on cycles, per independent cycle.

For a single cycle of $k$ edges, $|E| = k$, $b(G) = 0$, and $\mathcal{C}(G) = 1$, so $\Delta^{k}_{\beta}(G) = k - k = 0$.
As in the information-theoretic case, $\Delta^{k}_{\beta}$ vanishes on pure $k^{\textrm{th}}$-order interactions.

Fixing $k$ to be $0$, $1$, or $2$ gives graphical analogues of the S-information, dual total correlation, and (negative) O-information.

\subsubsection*{Graphical S-information ($\mathbf{k=0}$)}

\begin{align}
    \sinfo_{\beta}(G) = \sum_{e \in E}\Big[\mathcal{C}(G) - \mathcal{C}(G^{-e})\Big] = |E| - b(G) \label{eq:s_beta}
\end{align}
The graphical S-information counts the edges that lie on at least one cycle: each such edge, when deleted, destroys exactly one independent cycle.
Like the standard S-information, $\sinfo_{\beta}(G)$ is non-negative, and it vanishes if and only if $G$ is a forest.

\subsubsection*{Graphical dual total correlation ($\mathbf{k=1}$)}

\begin{align}
    \dtc_{\beta}(G) = \sinfo_{\beta}(G) - \mathcal{C}(G) = |V| - c(G) - b(G) \label{eq:dtc_beta}
\end{align}
This is the graphical analogue of $\dtc(\xjoint) = \sinfo(\xjoint) - \tc(\xjoint)$.
The quantity $|V| - c(G)$ is the number of edges in a spanning forest of $G$ (the rank of the graphic matroid).
Every bridge belongs to every spanning forest, so $\dtc_{\beta}(G)$ counts the spanning-forest edges that also lie on cycles.
It is therefore non-negative.

\subsubsection*{Graphical (negative) O-information ($\mathbf{k=2}$)}

\begin{align}
    \oinfo_{\beta}(G) = 2\,\mathcal{C}(G) - \sinfo_{\beta}(G) = |E| - 2|V| + 2c(G) + b(G) \label{eq:o_beta}
\end{align}
A single cycle of $k\geq3$ edges has $\oinfo_{\beta}=2-k<0$: isolated cycles are synergistic.
In dense graphs, however, cycles overlap and share edges, and $\oinfo_{\beta}$ can become positive (e.g. $\oinfo_{\beta}(K_5)=2$), indicating redundancy.

\end{document}